# Revealing the Origin of Time-reversal Symmetry Breaking in Fe-chalcogenide Superconductor FeTe$_{1-x}$Se$_x$

Camron Farhang[1], Nader Zaki[2], Jingyuan Wang[1], Genda Gu[2], Peter D. Johnson[2] and Jing Xia[1]

[1]*Department of Physics and Astronomy, University of California, Irvine, California 92697, USA*

[2]*Condensed Matter Physics and Materials Science Division (CMPMSD), Brookhaven National Laboratory, Upton, NY 11973*

Recently evidence has emerged in the topological superconductor Fe-chalcogenide FeTe$_{1-x}$Se$_x$ for time-reversal symmetry breaking (TRSB), the nature of which has strong implications on the Majorana zero modes (MZM) discovered in this system. It remains unclear however whether the TRSB resides in the topological surface state (TSS) or in the bulk, and whether it is due to an unconventional TRSB superconducting order parameter or an intertwined order. Here by performing in superconducting FeTe$_{1-x}$Se$_x$ crystals both surface-magneto-optic-Kerr effect (SMOKE) measurements using a Sagnac interferometer and bulk magnetic susceptibility measurements, we pinpoint the TRSB to the TSS, where we also detect a Dirac gap. Further, we observe surface TRSB in non-superconducting FeTe$_{1-x}$Se$_x$ of nominally identical composition, indicating that TRSB arises from an intertwined surface ferromagnetic (FM) order. The observed surface FM bears striking similarities to the two-dimensional (2D) FM found in 2D van der Waals crystals, and is highly sensitive to the exact chemical composition, thereby providing a means for optimizing the conditions for Majorana particles that are useful for robust quantum computing.

The quest for a robust quantum computer that is immune to external perturbations has stimulated intense searches for topologically protected quantum phases of matter where quasiparticles obey non-Abelian exchange rules [1]. One such example, the Majorana zero modes (MZM), have been reported in Fe-chalcogenide superconductors FeTe$_{1-x}$Se$_x$ as bound states in magnetic vortex cores [2–5] and as propagating 1D modes [6] by scanning tunnelling microscopy (STM) that performs spectroscopic imaging of the surface state. However, the imperfect conductance quantization [5] and the unexpected field dependence of MZM's occurrence [4] highlight the need for a deeper understanding of FeTe$_{1-x}$Se$_x$ towards the optimization of material conditions for MZM. The hallmark of topological superconductivity [7], the topological surface state (TSS) has been reported [8,9] in FeTe$_{1-x}$Se$_x$ by angle-resolved photo emission spectroscopy (ARPES) that probes the energy dispersions of surface electrons. Photoemission spectra reveal [8,9] a topologically protected TSS characterized by a linear dispersion centered at the Dirac point at chemical potential $\mu$ below $E_F$, and below the superconducting critical temperature $T_c$, a superconducting gap at the Fermi energy $E_F$. It is this superconducting TSS, when subjected to a magnetic field, that hosts the reported MZM that is potentially useful for topologically protected quantum computing. Without the magnetic field, the TSS is expected to obey time-reversal symmetry (TRS) and remain massless.

Therefore, it came as a surprise when in a low temperature ARPES study at zero magnetic field [10] the above linear dispersion was interrupted by the opening of a Dirac gap as temperature is lowered across $T_c$ in FeTe$_{1-x}$Se$_x$. The associated mass acquisition points to symmetry breaking. To explain this observation, a phenomenological Weiss field $h$ was introduced to the surface Hamiltonian, which fits well to the observed photoemission spectra [10]. A Weiss field in the absence of an external magnetic field represents spontaneous time-reversal symmetry breaking (TRSB), which could arise from ferromagnetism (FM) or an intrinsic TRSB superconducting (SC) state [11]. So far, the results of magnetic characterizations of FeTe$_{1-x}$Se$_x$ remain mixed. While nitrogen vacancy center (NV) magnetometry has detected static magnetic fields that are best described by a combination of SC supercurrents and FM in micron-sized exfoliated flakes [12], high resolution magnetic neutron scattering measurements of bulk crystals have revealed instead antiferromagnetic (AFM) orders of either double or single stripe spin arrangements [13]. Since ARPES is surface sensitive while NV magnetometry and neutron scattering probe predominantly the bulk, a critical step to solve this mystery is to separately identify the TRSB properties of the bulk and surface in FeTe$_{1-x}$Se$_x$. Another central issue is to resolve the relation between the possible TRSB and the SC order: is TRSB from a FM phase that competes with the SC; or does the SC state in FeTe$_{1-x}$Se$_x$ have an intrinsically TRSB order parameter analogous to the superfluidity [14] in $He^3$? If it is the first case, one might be able to separately control the FM and SC orders in the surface and engineer the optimal conditions for MZM in vortex cores. And the second scenario points to a new unconventional SC state. Based on ARPES spectra taken at discrete temperatures [10], the Dirac gap in TSS opens at a temperature close to $T_c$. Therefore the second scenario was favored in prior theoretical treatments [11], but needs to be tested by stringent experiments.

Surface-magneto-optic-Kerr effect (SMOKE) [15,16] measurements performed by a zero-area loop fiber optic Sagnac interferometer [17] are ideally suited for addressing these questions. In a conventional SMOKE setup, a linearly polarized light beam interacts with the surface magnetic moment $M$ through spin-orbit coupling and will experience a rotation $\theta_K$ of the polarization plane [18]. The Kerr rotation

$\theta_K$ is proportional to $M$, and thus provides a direct measurement of surface magnetization within the optical penetration depth $\delta$, typically a few nanometers for conductors [15,16]. For even higher resolution, we have introduced a zero-area loop [17] fiber optic Sagnac interferometer [19] that measures directly the non-reciprocal phase difference between counter-propagating circularly polarized light beams. It fundamentally rejects polarization rotations due to non-TRSB effects such as linear and circular dichroism [20].

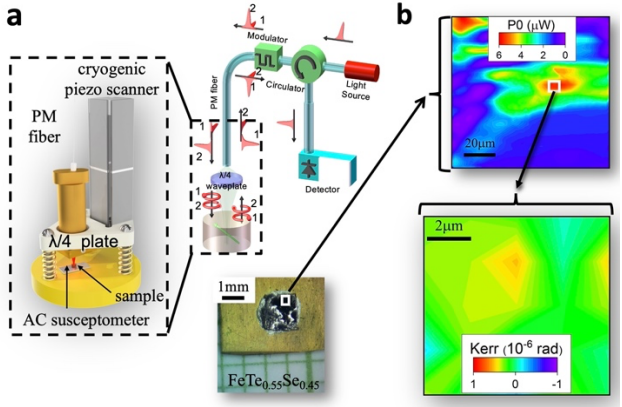

Fig. 1. (a) Sagnac setup for polar Kerr measurements (b) Scanning images of reflected optical power $P_0$ and Kerr signal ($\theta_K$) of a FeTe$_{0.55}$Se$_{0.45}$ sample (type B) at $1.8\ K$ and zero magnetic field.

As illustrated in Fig. 1a, for this study we utilize a scanning Sagnac microscope with $2\ \mu m$ lateral spatial resolution [21,22]. inside a cryostat with $1.8\ K$ base temperature and $9\ T$ magnetic field capability. The interferometer itself is maintained at room temperature. And a polarization maintaining fiber delivers light beams of orthogonal linear polarizations into the high vacuum sample space inside the cryostat. A cryogenic quarter wave ($\lambda/4$) plate converts the polarization of these light beams to circular polarization of opposite chirality that will interact with the sample surface and detect TRSB in the form of the non-reciprocal phase difference $\phi_{nr} = 2\ \theta_K$ when they finish the Sagnac loop and interfere at the detector. The Kerr resolution is at the ten nanoradian ($nrad$) level [23,24] that is about a thousand times better than conventional SMOKE [15,16]. The ARPES studies were carried out using the frequency quadrupled output of a 3-ps pulse width, 76-MHz repetition rate Coherent Mira 900P Ti:sapphire laser. The photoemission spectra were obtained using a Scienta SES 2002 electron spectrometer with an effective energy resolution of 2.5 meV. Bulk magnetic characterizations are performed using an AC susceptometer. Instrumentation details are presented in the Supplementary Information (FIG. S1-S3).

Single crystals of FeTe$_{1-x}$Se$_x$ with nominal x values of 0.3 and 0.45 were grown by a unidirectional solidification method (Supplementary Information FIG. S5). Flat $ab$ plane surface regions of tens of microns in size can be found (Fig. 1a inset) for measurements. We locate such flat regions with uniform reflected optical power ($P_0$) before performing spatial Kerr ($\theta_K$) scans at a fixed temperature or temperature scans at a fixed location. Fig. 1b demonstrates an example of this experimental procedure for a FeTe$_{0.55}$Se$_{0.45}$ sample. In the flat region inside the white square with a uniform $P_0 = 6\ \mu W$, the Kerr scan shows a signal of up to $\theta_K = 500\ nrad$ at the base temperature of $1.8\ K$ and zero magnetic field, indicating spontaneous TRSB. As is typical of spontaneous symmetry breaking, the sign and size of $\theta_K$ at zero magnetic field normally varies as a function of location and temperature, which agrees with the Kerr scan in Fig. 1b.

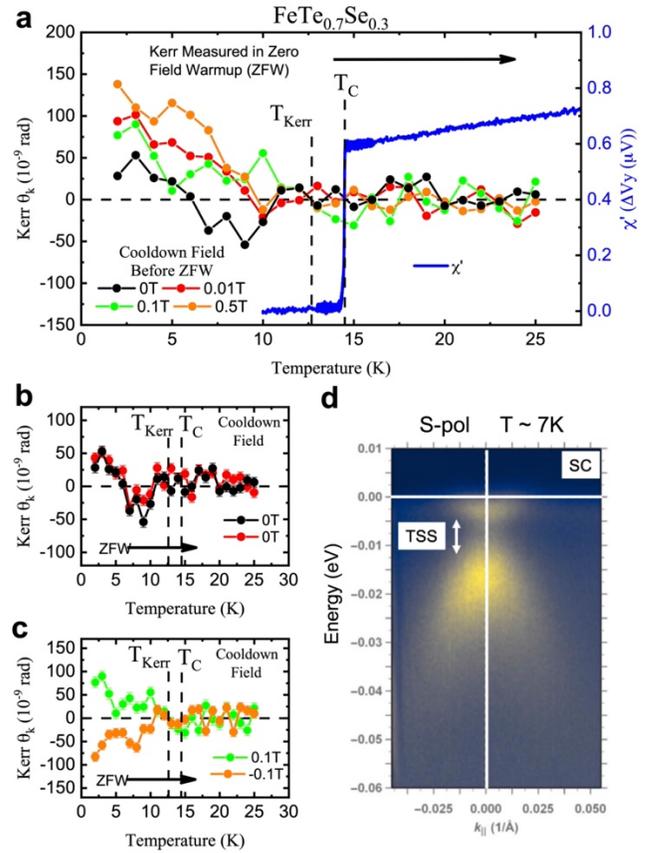

Fig. 2. Superconducting FeTe$_{0.7}$Se$_{0.3}$ (a) Kerr $\theta_K$ (left axis) up to $150\ nrad$ during zero field warmups (ZFW) indicate onset of spontaneous TRSB at $T_{Kerr}$; bulk magnetic susceptibility $\chi'$ (right axis) indicates onset of SC at $T_C$ without any sign of bulk FM. (b) $\theta_K$ during ZFW after zero field cooldown. (c) $\theta_K$ during ZFW after +/- 0.1T trainings during cooldown. (d) ARPES spectral intensity measured in the vicinity of the $\Gamma$-point ($k_{//} = 0$), using s-polarized light and with the sample in the superconducting state at $7\ K$.

To pinpoint the location of the observed TRSB, we first focus on a FeTe$_{0.7}$Se$_{0.3}$ sample that was studied by ARPES

in the prior publication [10] where a Dirac gap opens below the superconducting critical temperature $T_c = 14\ K$. This Dirac gap in the TSS can be seen in the 7 K ARPES spectral intensity map with s-polarized light in Fig. 2d. Bulk superconductivity is verified in AC susceptibility measurements (right axis, blue line) shown in Fig. 2a, where $\chi'$ displays a pronounced sharp diamagnetic Meisner drop when the sample is cooled below $T_c = 14\ K$. There is no sign in $\chi'(T)$ of any bulk FM transition, which agrees with magnetic neutron scattering [13]. In contrast, the surface Kerr signal measured during zero magnetic field warmups (ZFW) after zero magnetic field cooling (ZFC), as shown in Fig. 2b, display clear onsets of $\theta_K$ at $T_{Kerr} = 12.5\ K$. This serves as direct evidence of TRSB on the surface, either due to the formation of surface FM, or an unconventional TRSB SC state. As explained earlier, the sign and size of spontaneous $\theta_K$ are normally not fixed during ZFW after ZFC. Indeed in Fig. 2b polar Kerr signals $\theta_K$ of up to $\pm 50\ nrad$ have their sign fluctuating between positive and negative. In exfoliated flakes with a large surface-to-volume ratio, the TRSB surface might generate a magnetic field that is comparable in size to that from bulk SC, and account for the detected static field by nitrogen vacancy center (NV) magnetometry [12]. The AC susceptibility and surface Kerr measurements are consistent with ARPES spectra taken at 7 K (Fig. 2d) that shows the coexistence of the SC gap at $E_F = 0\ meV$ and the Dirac gap at the Dirac point of $-0.008\ eV$. We note that $T_{Kerr}$ and $T_C$ are close to each other within $1.5\ K$, and are difficult to distinguish in prior experiments [10,13]. We shall describe a decisive experiment later in this paper to tell them apart.

The sign of $\theta_K$ can be trained by cooling down in a symmetry-breaking magnetic field $B_{Cooling}$. This is demonstrated in Fig. 2c where $B_{Cooling} = \pm 0.1\ T$ during cooldowns result in $\pm 100\ nrad$ of spontaneous Kerr signal in subsequent ZFWs. In the case of FM, this is the well-known alignment of FM domains by an external magnetic field [25]. This training effect has also been demonstrated in TRSB superconductors [23,24,26,27], but with an important caveat. In a type II superconductor such as FeTe$_{1-x}$Se$_x$, magnetic vortices form when $B_{Cooling}$ is larger than the lower critical field $H_{C1}$ and penetrates the superconductor. After removal of $B_{Cooling}$, a small fraction of vortices can still be trapped at pinning sites, resulting in Kerr signals during ZFW. In fact, we speculate that motions of trapped vortices under thermal gradients may account for the reported spontaneous Nernst signal [28] in FeTe$_{1-x}$Se$_x$. Since $H_{C1}$ of $4\ mT$ [29] is smaller than the coercive field, in this sample trainings involve trapped vortices. To tell whether their contributions dominate $\theta_K$, Kerr signals were measured during ZFW after training fields of 0.01, 0.1 and 0.5 T as shown in Fig. 2a. We expect the number of trapped vortices hence their contributions to $\theta_K$ to roughly scale with training fields. However, remanent $\theta_K$ at $T = 1.8\ K$ only differ between 80 to 140 $nrad$ despite the 50 times difference in training fields, indicating that the trained $\theta_K$ in this sample are not dominated by trapped vortices.

Now we have experimentally established that TRSB occurs only on the surface, we turn to experiments on samples with the nominal chemical composition of FeTe$_{0.55}$Se$_{0.45}$ to identify the origin of this surface TRSB. Recently neutron scattering, ARPES, and resistivity measurements have been carried out to establish that FeTe$_{0.55}$Se$_{0.45}$ is located very close to phase boundaries in a complex phase diagram [13]. By a few percent change of the Fe concentration, at $T = 0$ a nominal FeTe$_{0.55}$Se$_{0.45}$ can be a non-superconductor (Type A), a SC with TSS (Type B), or a trivial SC without TSS [13]. In a Type A sample, neutron scattering has revealed a double stripe spin arrangement for the bulk AFM order, and the APRES spectra are featureless; while in a Type B sample, a single stripe spin arrangement is found, and the APRES spectra show both SC and TSS states.

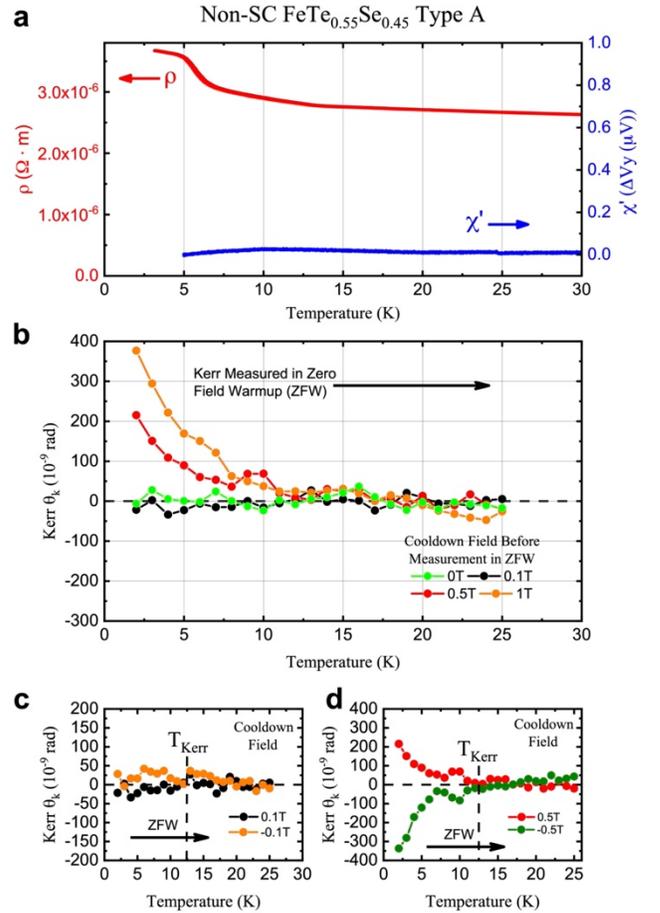

Fig. 3. Non-superconducting FeTe$_{0.55}$Se$_{0.45}$ (Type A) (a) Resistivity $\rho$ and bulk magnetic susceptibility $\chi'$ show no sign of SC or bulk FM transitions; (b) $\theta_K$ up to 400 $nrad$ during ZFW indicates onset of surface FM at $T_{Kerr}$; (c) $\theta_K$ during ZFW after $\pm 0.1\ T$ trainings. (d) $\theta_K$ during ZFW after $\pm 0.5\ T$ trainings.

We have performed measurements on both types of FeTe$_{0.55}$Se$_{0.45}$ samples. The experimental results on a Non-

SC (Type A) FeTe$_{0.55}$Se$_{0.45}$ sample are summarized in Fig. 3, where the ARPES spectra are featureless indicating the absence of TSS. And those on a SC+TSS (Type B) FeTe$_{0.55}$Se$_{0.45}$ sample are summarized in Fig. 4, with a Dirac gap in TSS and a SC gap at $E_F$ in the ARPES spectra (Fig. 4d) at $T \sim 5\ K$.

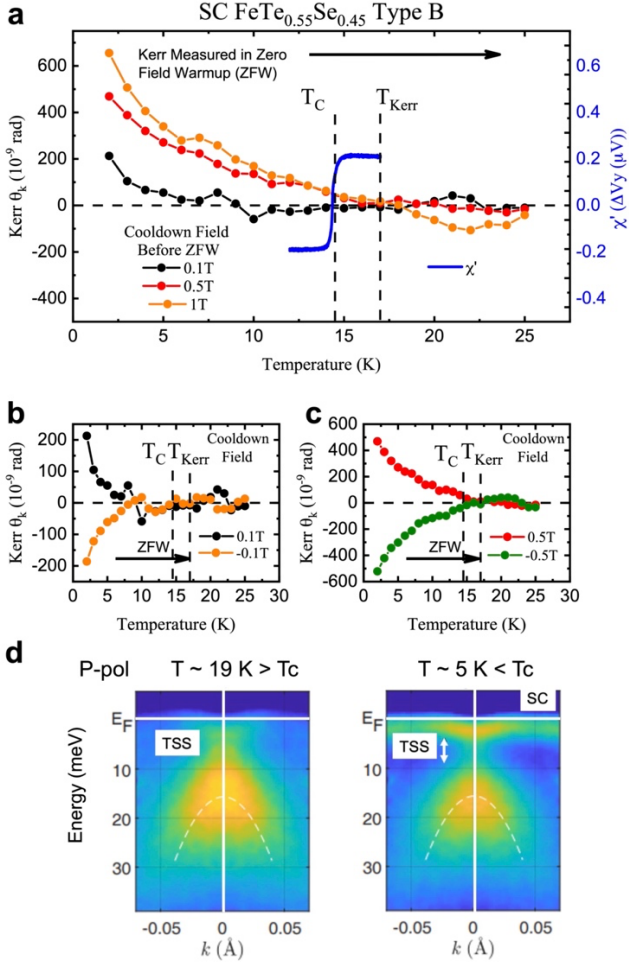

Fig. 4. Superconducting FeTe$_{0.55}$Se$_{0.45}$ (Type B) (a) $\theta_K$ (left axis) up to $600\ nrad$ during ZFW indicates onset of TRSB at $T_{Kerr}$; bulk magnetic susceptibility $\chi'$ (right axis) indicates onset of SC at $T_C$ without any sign of bulk FM. (b) $\theta_K$ during ZFW after $\pm 0.1\ T$ trainings. (c) $\theta_K$ during ZFW after $\pm 0.5\ T$ trainings. (d) ARPES spectral intensity measured in the vicinity of the Γ-point ($k_{//} = 0$), using p-polarized light, showing TSS at all temperatures, and the opening of a SC gap below $T_C$.

As shown in Fig. 3a, in the non-SC type A sample, the resistivity $\rho$ shows no sign of SC either in the bulk or on the surface. Magnetic susceptibility $\chi'$ confirms that there is no bulk SC, and it shows no sign of any bulk FM transition. Comparing the size of $\chi'$ in this non-SC sample to the drop of $\chi'$ across $T_C$ in the SC type B sample (Fig. 4a), we can estimate an upper bound of 1% for superconducting volume fraction, indicating that this type A sample is indeed deep in the non-SC region of the phase diagram [13]. Therefore, the onsets of the surface Kerr signal up to $\theta_K = 400\ nrad$ (Fig. 3b) during ZFW undoubtably indicate that the observed surface TRSB is not due to a TRSB order parameter of the SC state. Instead, it originates from surface FM. In a non-SC sample (Fig. 3), $T_{Kerr} > 0$ and $T_C = 0$; while in superconducting samples (Fig. 2 and Fig. 4), $T_{Kerr}$ and $T_C$ are not necessarily identical, but they are close to each other. The observed closeness between $T_{Kerr}$ and $T_C$, being a mere coincidence or not, puts the competing FM and SC orders very close in energy, and may be partially responsible for the complex phase diagram [13].

Note that there is no contribution to $\theta_K$ from trapped vortices for the entire temperature range in the Type A sample, and for $T_C < T < T_{Kerr}$ in the Type B sample. With the absence of any bulk FM signal, these two samples further confirm surface localization of TRSB.

The temperature dependences of the spontaneous Kerr signal $\theta_K$ (Fig. 2a, Fig. 3a, and Fig. 4a) in FeTe$_{1-x}$Se$_x$ bare striking similarities to the two-dimensional (2D) ferromagnetism in 2D van der Waals (vdW) crystals of Cr$_2$Ge$_2$Te$_6$, especially the bilayer case [21] (Supplementary Information FIG. S4). Namely, $\theta_K(T)$ doesn't saturate quickly with a reduced temperature following the well-known $\tanh\left(\frac{\mu B_E}{k_B T}\right)$ function of the 3D Heisenberg model [25], where $B_E$ is the molecular field. Instead, like in the exfoliated bilayers of Cr$_2$Ge$_2$Te$_6$, $\theta_K(T)$ appears to keep growing at the lowest temperatures [21], which is a direct consequence of the fact that 2D magnetism is stabilized by magnetic anisotropy instead of magnetic exchange coupling [21]. The surface FM in FeTe$_{1-x}$Se$_x$ is sensitive to the chemical composition, indeed, the amplitude of observed spontaneous $\theta_K$ in FeTe$_{0.55}$Se$_{0.45}$ of both type A and B are four times larger than in FeTe$_{0.7}$Se$_{0.3}$. In addition, while a training field of a mere $0.01\ T$ is enough to achieve a saturating $\theta_K$ in FeTe$_{0.7}$Se$_{0.3}$, $0.5\ T$ is needed in FeTe$_{0.55}$Se$_{0.45}$ samples. The fifty times difference in coercivity suggests a large difference in the magnetic anisotropy between FeTe$_{0.55}$Se$_{0.45}$ and FeTe$_{0.7}$Se$_{0.3}$. In contrast, the SC and FM onset temperatures $T_C$ and $T_{Kerr}$ only shift by a few Kelvins between FeTe$_{0.7}$Se$_{0.3}$ and SC FeTe$_{0.55}$Se$_{0.45}$. While it is premature to speculate on the origin of the surface FM, we note that magnetism could occur at surfaces and interfaces where inversion symmetry is broken [30].

In summary, we have unambiguously identified TRSB in FeTe$_{1-x}$Se$_x$ and pinpoint its origin to be surface ferromagnetism. The detected polar Kerr signals at zero magnetic fields indicate a perpendicular component of the surface magnetization, which is necessary to form the observed Dirac gap. The surface FM is highly sensitive to the exact chemical composition, which would allow exploring the optimal conditions for stabilizing MZM in magnetic vortex cores [2–5] and 1D modes [6] useful for quantum computing [1].

We acknowledge discussions with C. Wu and I. Zaliznyak. This project was supported mainly by NSF award DMR-1807817, and in part by the Gordon and Betty Moore Foundation through Grant GBMF10276. The work at BNL was supported by the US Department of Energy, office of Basic Energy Sciences, contract no. DOE-sc0012704.

email: xia.jing@uci.edu